\documentclass[aps,twocolumn,twoside,floatfix,nofootinbib,a4paper]{revtex4}
\usepackage{graphicx}
\usepackage{marginnote}
\usepackage{bbm}
\usepackage{hyperref}
\usepackage{enumitem,kantlipsum}
\usepackage[usenames,dvipsnames]{color}
\usepackage{amsmath}
\usepackage{amssymb}
\usepackage{amsfonts}
\usepackage{mathrsfs}
\usepackage{bm}
\usepackage[all]{xy}
\usepackage{hyperref}
\usepackage{tikz-cd}

\newcommand{\beq}{\begin{equation}}
\newcommand{\eeq}{\end{equation}}
\newcommand{\bea}{\begin{eqnarray}}
\newcommand{\eea}{\end{eqnarray}}
	
\newcommand{\ea}{\end{align*}}

\newcommand{\bma}{\begin{pmatrix}}
\newcommand{\ema}{\end{pmatrix}}

\usepackage[percent]{overpic}
\usepackage{overpic}

\begin{document}
\title{Radio filaments as Z-pinched Galactic center wind}
\author{Fan Zhang} 
\affiliation{Institute for Frontiers in Astronomy and Astrophysics, Beijing Normal University, Beijing 102206, China}
\affiliation{Gravitational Wave and Cosmology Laboratory, School of Physics and Astronomy, Beijing Normal University, Beijing 100875, China}
\affiliation{Advanced Institute of Natural Sciences, Beijing Normal University at Zhuhai, Zhuhai 519087, China}

\date{\today}

\begin{abstract}
\begin{center}
\begin{minipage}[c]{0.9\textwidth}
In this brief note, we tentatively investigate the possibility that the radio filaments are produced when the Galactic center wind washes over magnetic field structures. The electrons and ions, with their disparate charge-to-mass ratios, are deflected differently by the magnetic field, and a current results. The current is subsequently Z-pinched into filaments, creating an electron-accelerating electric field along the way, because the magnetic field necessarily rearranges during the dynamic constriction process. An axial magnetic field also arises, possibly via the diocotron channel, to eventually quench the pinching and stabilize the filaments against a variety of instabilities. 
\end{minipage} 
 \end{center}
  \end{abstract}
\maketitle

\raggedbottom

\section{Introduction \label{sec:intro}}
The cause of the non-thermal\footnote{Some shorter thermal filaments may constitute a separate population, see e.g., \cite{2023ApJ...949L..31Y}. We are concerned with the longer non-thermal filaments in this note, but some aspects of our discussion may nevertheless be portable to those other populations.} radio filaments permeating the inner couple of degrees of the Galactic center \cite{1984Natur.310..557Y,1985ApJ...293L..65L,1986ApJ...310..689Y,1991Natur.353..237G,1992A&A...264..500H,1999ApJ...526..727L,2004ApJ...607..302L} (see also e.g., \cite{2004ApJS..155..421Y,2008ApJS..177..515L,2022ApJ...925L..18Y,2022MNRAS.515.3059Y,2022MNRAS.517..294Y,2022ApJ...939L..21Y} for recent population surveys), oriented roughly perpendicular to the Galactic plane, has remained elusive, but see e.g., \cite{1992A&A...264..493L,1996ApJ...470L..49R,1995ApJ...443..638N,1999ApJ...521..587S,2001ApJ...548L..69B,2002ApJ...568..220D,2003ApJ...598..325Y,2006ApJ...637L.101B,2011ApJ...741...95L,2016MNRAS.455.1309B,2019MNRAS.490L...1Y,2020ApJ...890L..18T,2020PASJ...72L...4S,2021MNRAS.501.1868C,2023MNRAS.518.6273S} for some proposals. In this note, we put forward yet another hypothesis, caricatured as follows: 
\begin{enumerate}
\item[A:] It had been suggested by e.g., \cite{2019MNRAS.490L...1Y,1999ApJ...521..587S,2000ApJ...543..775G,2002ApJ...568..220D}, that a Galactic center wind (see e.g., \cite{2011MNRAS.411L..11C} for observational support, and also \cite{2013ApJ...762...33Y,2019ApJ...883...54O} for the existence of hot gas in the Central Molecular Zone -- can be crudely demarcated and visualized as a slab sitting on the Galactic plane \cite{2019MNRAS.490L...1Y} -- that could expand into a wind) blowing outwards away from the Galactic plane creates the filaments after encountering obstacles such as stellar wind bubbles or molecular clouds, like how cometary tails form. We propose an alternative here, which assigns a different r\^ole for the obstacles, that entails a less intense interaction with the wind, and allows for a weaker spatial correlation against the filaments (see item B below and e.g., \cite{2022MNRAS.517..294Y} for indefinite but tentatively existent observational associations). 

Specifically, the plasma within the partially ionized wind (see e.g., \cite{2016A&A...585A.105L,2019ApJ...883...54O} for high ionization rate in the Central Molecular Zone sourcing the wind) will encounter local/fragmented intra-Galactic magnetic structures (e.g., those associated with closely-packed clusters of HII regions, stellar wind bubbles or individual molecular clouds; local inhomogeneities in the magnetic field not associated with any baryonic matter over-density may also be relevant), that are relatively numerous and strong within the Galactic center even off the plane (cf., the Galactic bulge that protrudes from the disk). The ions and electrons experience similar Lorentz force strengths when comoving, due to similar electric charge amplitudes, but would end up with quite different accelerations due to their vast mass differences (as well as their opposite charge signs). As a result, whatever the magnetic field configuration, electrons will be more strongly deflected away from the original vertically ascending (in Galactic latitude) trajectory than ions, and the bulk velocity differential between the species thus forged leads to a current. 

The current is subsequently shredded by the Z-pinching (see e.g., \cite{Haines_2011}; this is a plasma phenomenon utilized in early controlled-fusion attempts), resulting from the simple fact that parallel current elements attract each other according to the Biot-Savart force law \cite{1934PhRv...45..890B}, and ends up exhibiting a characteristic filamentary morphology\footnote{\label{fn:backbone}This shredding is ubiquitous in plasma physics contexts, giving rise to filamental morphologies commonly observed within all sorts of free-standing plasma \cite{2015ppu..book.....P,2002ApJ...568..220D,1992ITPS...20..867C}. There are indeed a few previous proposals for the explanation of radio filaments, that enlist pinching, see e.g., \cite{1988ApJ...333..735B,BATrubnikov_1990}. These differ from the present discussion in terms of the source of current and other core ingredients, such as what is the dominant expansive counter-force during steady-state. In particular, they rely on a highly-ordered ``backbone'' background poloidal magnetic field at the Galactic center (see e.g., \cite{2024ApJ...969..150P} for recent observational data that do not appear to exhibit such a background field outside of the filaments; if anything, the field is azimuthal rather than poloidal near the Center Molecular Zone \cite{2010ApJ...722L..23N}), while our axial magnetic field is self-generated and the large scale orderliness along the filament is shepherded by the Galactic center wind instead. A key consequence of this divergence is that unlike the magnetic field, which is a pseudo-vector, the wind-launching pressure gradient is a true vector, with odd parity, so the filaments it guides are less likely to develop helical shapes (with a consistent and definite handedness; a backbone bias magnetic field picks out a preferred handedness but pressure gradient does not, so such a strong dominance of one handedness over the other would have to rely on extreme small probability statistics). Indeed, unlike some other filamentary structures (see e.g., \cite{1992ITPS...20..867C}), the radio filaments don't seem to exhibit a spirally morphological trait (see e.g., \cite{2019ApJ...884..170P}).}.

\item[B:] The Z-pinching, as per the generalized Bennett relation \cite{1981PhRvA..24.2758W,1988Ap&SS.144...73C} (see item C below for more details), naturally produces an axial electric field. This is because as the azimuthal/toroidal (labelled with $\phi$ hereafter) magnetic field redistributes during the pinching, its temporal variation creates a toroidal magnetic displacement current, which equals to a curl of the electric field, consonant with an electrical circulation in the poloidal/axial ($z$) direction (cf., a familiar solenoid but with electric and magnetic fields swapped, and current replaced by displacement current). 

This electric field constitutes one of the expansive forces according to the generalized Bennett relation, that resist the magnetic constriction, i.e., it enacts an expected inductance. However, electric fields tend not to persist in astrophysical settings. Specifically, its energy gets drained when it accelerates the electrons into relativistic rapidities, capable of emitting the non-thermal synchrotron radiation when subjected to the ${\bf v}\times {\bf B} \sim v_z B_{\phi}$ force (Z-pinches giving off radio emission is a familiar phenomenon from both solar physics \cite{Takakura1960SynchrotronRF,Kawabata1964TransferOT,1965PASJ...17..294K} and laboratory experiments \cite{1984PhR...104..259M}). This eventual loss of energy into synchrotron radiation allows Z-pinching to continue. 

\item[C:] 
Although the compressive influence of the Z-pinching is executed by an azimuthal magnetic field, self-generated by an axial current, there can well also be an axial magnetic field being generated during the pinching. To see why this is plausible, recall first that the generalized Bennett relation, describing the radial redistribution of plasma matter density, contains both expansive and compressive terms. 
The compressive forces that could cause the pinching include the usual one due to the azimuthal magnetic field that already appears in the non-generalized Bennett relation \cite{1934PhRv...45..890B}, and a self-gravity of the plasma column (negligible for our context due to low density, see Eq.~\ref{eq:GravPressure} below). 
The expansive terms on the other hand, include a transverse electric field contribution due to charge separation\footnote{\label{fn:ChargeSep}The magnetic field deflection pulls electrons and ions in opposite directions due to their opposite charges, so this effect is relevant for us. However, like the axial electric field, this transverse field may eventually be eradicated by e.g., the development of diocotron instability, that charge separation tends to stimulate \cite{2015ppu..book.....P}.}, and a few other excess (in the filament as compared to environmental, with the latter represented by values recorded at the filament surface) energy terms associated with an axial magnetic, an axial electric field (see item B above), the plasma radial motion, the centrifugal force if the filament is rotating, and thermokinetics (this is the balancing term in the original Bennett relation, narrating outward thermal pressure). 

In order for the Z-pinching to reach a stabilized quasi-equilibrium state (so the filaments are not short-lived transients, and it is reasonable that we observe them in abundance), the expansive terms must grow to eventually match and balance the compressive force, likely through shifting some of the axial plasma-flow kinetic energy into one of the aforementioned excess energy terms \cite{2015ppu..book.....P}. The appearance of a significant inhomogeneous axial magnetic field is one of the options\footnote{Recall that from magnetic field energy considerations, one may very crudely intuit the magnetic field as being akin to an elastic body that resists compression in directions transverse to the field lines (and stretching along the lines). So when there is an appreciable axial magnetic field, its magnetic pressure can replace the usual thermal pressure in the Bennett pinch, to balance the constriction from the current-erected azimuthal magnetic field.} (the leaky axial electric field cannot fulfill this r\^ole), and an attractive one since it is well-known that the presence of an additional axial magnetic component (giving rise to magnetic shear) tends to stabilize the Z-pinching against some of the more structurally destructive large length-scale instabilities (see e.g., \cite{Suydam1958StabilityOA,1982RvMP...54..801F,1991SvPhU..34.1018T,2000RvMP...72..167R}; for a filament aspect ratio of 100, the axial field strength should exceed its azimuthal counterpart by at least an order of magnitude to stave off sausage and kink instabilities, as per Eqs.~20 and 21 in \cite{1988ApJ...333..735B}), thus also helps with the continued maintenance of the steady-state, after establishing it in the first place\footnote{One way to make sense of this is by noting that the presence of a strong axial magnetic component allows for the current flow and the magnetic field to become aligned, thus settle into a force-free configuration (see also comment below Eq.~2.53 of \cite{2015ppu..book.....P}). This is consistent with the general theoretical expectation, that the presence of shearing in the plasma flow, due to e.g., the diocotron instability, would give rise to a non-potential magnetic field, which precisely tends to settle into such a force-free configuration \cite{2015ppu..book.....P}. This configuration is the lowest in magnetic energy that can be attained in a closed system, so when the total plasma system is dissipative (e.g., via synchrotron radiation losses), it is a natural candidate for the steady-state.}. 

Indeed, observations (see e.g., \cite{1984PASJ...36..633I,1995PASJ...47..829T,1997ApJ...475L.119Y,2019ApJ...884..170P,2024arXiv240816745P}) confirm the presence of axial magnetic components in the filaments. In particular, because the compressive force is proportional to the total integration of the energy density of the azimuthal magnetic field (see e.g., Sec.~2.5 of \cite{2015ppu..book.....P}), while the expansive counterpart is an inhomogeneity energy instead, proportional to a similar integral for the axial field but minus the integration result when the field strength is held constant at the surface value (i.e., the expansive term vanishes if the axial field is radially constant), a stronger axial field component (than its azimuthal counterpart) would thus generally be needed to achieve a force balance. This is consistent with most filaments exhibiting a more axial magnetic field orientation\footnote{There are also a few examples where an azimuthal advantage reigns over parts of the filaments, which could possibly be due to complications in the local environment. For example, the entrainment or injection of additional slow (in the axial direction) electrons, from adjacent celestial bodies, would increase the ratio between the Budker's parameter and electron beam rapidity, so these filaments reach towards the Alfv\'en limiting current. To avoid exceeding the limiting current, the plasma motion could become less beam-like along the axial direction, and pick up more significant transverse and randomized components (see e.g., Sec.~2.9.2 of \cite{2015ppu..book.....P}). As a result, the force-balance may in this case be accomplished by the transverse motion or thermokinetics terms, instead of an axial magnetic field term. For a specific example where this mechanism might be at work, note that the region accommodating a predominantly transverse magnetic field in SNTF1 (see Fig.~7 in \cite{2024arXiv240816745P}) appears to coincide with where the outflow from a jetted point source intersects the filament (see Fig.~2 in \cite{2022ApJ...941..123P}). 
Alternatively of course, we could simply be witnessing earlier stages of pinching, before stabilization.}.

\end{enumerate}

\section{Practical portion \label{sec:Env}}
In this section, we plug in some real numbers related to the processes outlined in the Introduction Sec.~\ref{sec:intro}, thereby provide further information on their plausibility. The parameter values we assume are as follows, and we caution that each is subject to uncertainties or even disputes: 
\begin{itemize}
\item 
The axial magnetic field strength along filaments: $B^f_z\sim \mathcal{O}(10^{-1})-\mathcal{O}(10^0)\,$mG as per observations by e.g., \cite{2009A&A...505.1183F,2022MNRAS.515.3059Y}. 

\item 
The magnetic field strength in local structures: from $B \sim \mathcal{O}(10^1)\,{\rm\mu}$G in molecular clouds (see e.g., \cite{1999ApJ...515..275C}) to up to $B \sim \mathcal{O}(10^0)\,$mG in HII regions (see e.g., \cite{1981ApJ...247L..77H})\footnote{These do not significantly exceed $B\sim \mathcal{O}(0.1)\,$mG in the Central Molecular Zone (see e.g., \cite{2019ApJ...883...54O}) and perhaps the weaker field just outside (the equipartition magnetic field of the Galaxy as a whole is in the range of tens of $\mu$G), which suggests that some inhomogeneities in the magnetic field itself, not associated with any baryonic matter features, may also provide equally strong macroscopic scattering centers. In such cases, the resulting filaments will not have obvious associated celestial objects embedded upstream in the Galactic wind.}.  

\item 
The plasma density is assumed at $n_i \approx n_e \sim \mathcal{O}(10^{-2})-\mathcal{O}(10^{-1}) \,\text{cm}^{-3}$ in the Galactic center ($e$ stands for electrons and $i$ for ions, assumed to be mostly protons and some alpha particles, whose mass and charge diversities are neglected in our order-of-magnitude estimates; we also ignore charged dust particles here, as the magnetic fields involved are weak, thus dusts with minuscule charge-to-mass ratios can be treated as being approximately neutral).

\item The temperature of the Galactic center medium is taken to be at the $T \sim \mathcal{O}(10^2)\,$K rung (1eV $\approx 10^4\,$K), corresponding to the kinetic temperature of the warm gas (assumed to be in thermal equilibrium, so electron and ion temperatures are the same) in the Central Molecular Zone that supposedly launched the Galactic center wind, see e.g., \cite{2019ApJ...883...54O,2019MNRAS.490L...1Y}.   
\end{itemize}

With these parameters, the Debye length of the pre-pinching plasma (swapping to post-pinching parameters does not change the qualitative conclusions)
\bea
\lambda_D = \sqrt{\frac{\epsilon_0 k_B T}{n_e e^2}} 
\eea
is measured in mere meters, far smaller than the parsec scales of the astrophysical structures that we are investigating, thus it is legitimate to treat the charged particles as a collective plasma, rather than having to consider them individually. We will therefore liberally lift formulae from plasma physics literatures. For the rest of this section, the subsection indices correspond to those in the itemized list in the Introduction Sec.~\ref{sec:intro}.

\subsection{Charge discrimination\label{sec:Current}}
The first task is to create the current that ends up becoming pinched. The size of this current can be evaluated by noting that the final axial-magnetic-field-sustained quasi-equilibrium is achieved at a current of (Eq.~2.60 of \cite{2015ppu..book.....P}; $a$ is filament width; square brackets enclose the units, and SI units are assumed when they are absent, we switch between different unit systems sometimes to stick with formulae in their most convenient form)
\bea \label{eq:Current}
I[\text{A}]\approx (3\times 10^{12}) \times \frac{2 \pi a[\text{pc}] \, B^f_z[\text{G}] }{\mu_0}\,,
\eea
which translates into a mere 
\bea
\Delta v_{\bot} \sim \mathcal{O}(10^{-4})-\mathcal{O}(10^{-2}) \,\text{m/s}
\eea
vertical bulk speed difference between the electrons and ions, when the width of the filaments is taken to be around the typical $a \sim 0.5\,$pc \cite{2007astro.ph..1050M,2022ApJ...939L..21Y}. 
This is much smaller than the thermal speeds\footnote{Thus we do not suffer the Buneman/two-stream instability \cite{1959PhRv..115..503B}, and cannot invoke double layers \cite{1977dlr..rept.....B} as an electron acceleration mechanism (cf., \cite{BATrubnikov_1990}).}
\bea \label{eq:ThermalSpeedVals}
\bar{v}_{e} \sim \mathcal{O}(10^5)\,\text{m/s}\,, \quad \bar{v}_{i} \sim \mathcal{O}(10^3)\,\text{m/s}\,, 
\eea
computed according to 
\bea \label{eq:ThermalSpeed} 
\bar{v}_{\alpha} =\sqrt{\frac{8k_B T}{\pi m_{\alpha}}}\,, \quad \alpha \in \{e, i\}\,.
\eea

The collective bulk motion wind speed $v_{\rm wind}$ is also much higher, at $\mathcal{O}(10^5)-\mathcal{O}(10^6)\, \text{m/s}$ \cite{2011MNRAS.411L..11C}, and the Lorentz force $\sim e v_{\rm wind} B$ experienced by the plasma particles traveling with the wind, traversing the magnetic structures, would give rise to accelerations on the order of $\mathcal{O}(10^7)-\mathcal{O}(10^{10})\, \text{m/s}^2$ for electrons and three orders of magnitudes smaller for ions. The magnetic field will deflect electron (and ion) bulk motions according to geometric arrangements, and cancelations are likely as the field needs to close up into loops. Collisions and other hydrodynamic processes may also resist further differentiation. Nevertheless, our quick appraisal above indicates that such field strengths as listed at the beginning of Sec.~\ref{sec:Env}, appear powerful enough to achieve sufficient efficacy so the desired $\Delta v_{\bot}$ be produced\footnote{Note this discussion refers to the pre-pinching phase. Once Z-pinching sets off, conversion into an axial magnetic field dominance, or in other words, a force-free configuration, occurs (see Sec.~\ref{sec:axialmag} below). As the name suggests, this would turn off the Lorentz forces and shutdown the differential acceleration presently being considered.}. Due to the convoluted nature of the many physical processes involved, and the lack of observational certainty regarding the wind and interloping magnetic field specifics however, we are unable to attempt a detailed quantitative analysis of the deflection process post-haste (but cf., the Hall effect for an analogy in controlled settings).

Nevertheless, we can bluntly assess whether some of the more obvious candidates could suppress the deflection mechanism. Firstly, we compute the Hall parameter 
\bea
h_{\alpha} = \frac{\omega_{\rm gyro \, \alpha}}{\nu_\alpha} \,, 
\eea 
i.e., gyrofrequency when traversing magnetic structures
\bea
\omega_{\rm gyro \, \alpha} = \frac{e B}{m_{\alpha}}\,,
\eea
over collision frequency 
\bea \label{eq:collfreq}
\nu_{\alpha} = \frac{\bar{v}_{\alpha}}{\lambda_{\alpha}} \,,
\eea
while the mean free paths are given by \cite{7169570}
\bea
\lambda_{\alpha}[{\rm pc}] \approx \frac{(T[\text{keV}])^2}{n_{\alpha}[\text{cm}^{-3}]}\,. 
\eea
Plugging in the numbers from the beginning of Sec.~\ref{sec:Env}, we obtain 
\begin{align} \label{eq:SegVals}
\lambda_{\alpha} \sim&\, \mathcal{O}(10^7)-\mathcal{O}(10^8)\,\text{m}\,, \notag \\
\nu_{e} \sim&\, \mathcal{O}(10^{-4})-\mathcal{O}(10^{-3})\,\text{s}^{-1}\,, \notag \\ 
\nu_{i} \sim&\, \mathcal{O}(10^{-6})-\mathcal{O}(10^{-5})\,\text{s}^{-1}\,,
\notag \\
\omega_{\text{gyro}\, e} \sim&\, \mathcal{O}(10^2)-\mathcal{O}(10^{4})\,\text{s}^{-1}\,,
\notag \\
\omega_{\text{gyro}\, i} \sim&\, \mathcal{O}(10^{-1})-\mathcal{O}(10^{1})\,\text{s}^{-1}\,,
\notag \\
h_{e} \sim&\, \mathcal{O}(10^5)-\mathcal{O}(10^{8})\,, \notag \\
h_{i} \sim&\, \mathcal{O}(10^4)-\mathcal{O}(10^{7})\,. 
\end{align}
Therefore, within our context, although the magnetic field is rather weak, the plasma is even more tenuous, so the Hall parameter values can still be quite large, indicating that the magnetic-field-deflection-induced velocity differential between species would not be quickly undone by the homogenizing effect of particle collisions. 

One could also compute the plasma $\beta$ parameter (ratio between kinetic -- due to transport of excess momentum, may be present even when collisionless -- and magnetic pressures), given by \cite{7169570}
\bea \label{eq:WindBeta}
\beta = (4\times 10^{-11})\times \frac{n_e[\text{cm}^{-3}] \left(T_e [\text{eV}] + T_i[\text{eV}]\right)}{B[\text{G}]^2}\,,
\eea
which ranges between $\mathcal{O}(10^{-4})-\mathcal{O}(10^{-7})$. In other words, while traversing the magnetic structures, the ``magneto'' likely dominates over the intrinsic wind ``hydrodynamics'', in terms of determining the overall magnetohydrodynamic behaviour of the plasma particles, so deflections should not be undone by kinetic transport effects.  
  
Finally, we notice it is encouraging that the total current from Eq.~\eqref{eq:Current} crops up at $I\sim \mathcal{O}(10^{15})-\mathcal{O}(10^{16})\,$A, which sits suitably in-between peers found within smaller structures such as the Orion nebula, at $\mathcal{O}(10^{14})\,$A (see Sec.~2.6.4 of \cite{2015ppu..book.....P}), and larger structures such as the Galactic whole, at $\mathcal{O}(10^{19})\,$A (see Sec.~2.6.5 of \cite{2015ppu..book.....P}), rather than coming out at some absurd number.
Substituting these values into dimensional analysis expressions (see e.g., \cite{2015ppu..book.....P}), we obtain that the compressive pressure due to the filaments' self-generated azimuthal magnetic field lands on 
\bea \label{eq:Pressure}
P_{\rm in} = \mu_0 \left(\frac{I}{a} \right)^2 \sim \mathcal{O}(10^{-9})-\mathcal{O}(10^{-7}) \, \text{Pa}\,.
\eea
Compare this with the gravitational self-attraction 
\bea \label{eq:GravPressure}
G m_i^2 (n_i a)^2 \sim \mathcal{O}(10^{-24})-\mathcal{O}(10^{-22}) \, \text{Pa}\,,
\eea
we see that it is justified to ignore the latter as we have done in Sec.~\ref{sec:intro}, and the pinching would indeed be due to the current computed in this section.    
  
\subsection{Electron acceleration \label{sec:Seg}}
For electrons to become relativistic and emit synchrotron radiation, its Lorentz factor $\gamma$ should be noticeably different from unity, i.e., at least the kinetic energy should be comparable to the rest mass energy, at $\sim\,$MeV. Getting this $V_{\rm min} \sim 10^6\,$V across the vast distance of tens to hundreds of parsecs (we set $L \sim 100\,$pc below as a typical value), we only need an electric field on the order of $E_{\rm min}\sim 10^{-13}\,$V/m. Substituting this strength into the Maxwell's equation (boldface denotes vectors)
\bea \label{eq:induction}
\frac{E_{\rm min}}{L} \sim \left| \nabla \times {\bf E} \right|= \left|\frac{\partial {\bf B}}{\partial t} \right| \sim \frac{B_{\phi \text{ min}}}{\tau}\,, 
\eea
we can gauge the minimal required azimuthal magnetic field strength $B_{\phi \text{ min}}$, and see if it is available within the context of our proposed pinching process. 

To this end, we need to compute the pinching timescale $\tau$, by substituting in the pressure from Eq.~\eqref{eq:Pressure} to obtain the radial acceleration at ($w$ here represents the pre-pinching width of the filament)
\bea
{\rm accl} \sim \frac{P_{\rm in}}{n_e M_i w}\,,
\eea
and noting on the other hand
\bea
w \sim \frac{1}{2}{\rm accl} \, \tau^2\,,
\eea
to solve for $\tau \sim \mathcal{O}(10^2)-\mathcal{O}(10^3)\,$yr. The lower end of this na\"ively guesstimate range is longer than the time it takes light to traverse the tens of parsecs if $w$ is set by the linear dimension of molecular clouds. It thus conveys the impression that the contraction could well be quite rapid, yet not unachievable. 

Plugging this $\tau$ into Eq.~\eqref{eq:induction}, we find that $B_{\phi \text{ min}} \sim \mathcal{O}(10^{-17})\, $G. For comparison, we can plug in the current computed in Sec.~\ref{sec:Current} into Amp\`ere's law to obtain the intrinsic self-generated azimuthal magnetic field
\bea
B_{\text{Amp } \phi} \approx \frac{\mu_0 I}{2\pi a} \sim \mathcal{O}(10^{-1})-\mathcal{O}(10^{0})\, \text{mG}\,.
\eea
This is not the only available source for $B_{\phi}$ however, since the extrinsic magnetic field, from the interloping magnetic structure, will also be affected by the pinching, because the magnetic Reynold's number for the plasma, given by ($\mathcal{M}$ is the Mach number) \cite{7169570}
\bea
\text{Re}_{\text{M}} \approx \mathcal{M} L[\text{cm}] (T[\text{eV}])^2\,,
\eea
is quite large, so the development of the magnetic field will not be a simple diffusion away of the inhomogeneities (i.e., resistive dissipation is not strong), but instead would be convected by the pinching flow. Both sources of the azimuthal magnetic field far exceed the minimum floor laid down, so there is ample room for the electron Lorentz factor to be boosted to much greater values, and the acceleration process can manage with much lower efficiency (e.g., some of the energy may be appropriated by diocotron instability, see Sec.~\ref{sec:axialmag} below). 

To ensure that collision-induced dynamical friction does not frustrate the acceleration by the pinching-produced electric field, we should also check the critical speed. Specifically, once an axial electric field is present, there would always be runaway (see e.g., Sec.~4.6.4 in \cite{2015ppu..book.....P}) electrons that keep getting accelerated, but these are only the electrons initially traveling faster than the critical value of 
\bea
v_{\rm crit} = \sqrt{\frac{e^3 n_e \ln\Lambda}{2\pi \epsilon_0^2 m_e E}}\,,
\eea 
which is only weakly dependent on temperature via 
\bea
\Lambda = \frac{12\pi }{n_e^{1/2}} \left(\frac{\epsilon_0 k_B T}{e^2}\right)^{3/2}\,.
\eea
Plugging in the numbers with $E=E_{\rm min}$ determined above, we obtain $v_{\rm crit} \sim \mathcal{O}(10^6)-\mathcal{O}(10^7)\,$m/s, which is comparable to the electron thermal speed given in Eq.~\eqref{eq:ThermalSpeedVals}. In other words, a large fraction of all electrons (and not some extreme tip of the statistical tail) will become runaway and be accelerated into relativistic rapidities. This high proportion of effective emitters partially\footnote{Another reason is that these emitters are also largely coherent, as we have an orderly electron flow, accelerated by a large-scale (homogeneous across volumes corresponding to individual pixels of the telescopes) electric field, impinging on a large-scale magnetic field.} explains why the tenuous plasma could possibly shine so brightly in radio. Specifically, substituting $L\sim 100\,$pc, $w\sim 10\,$pc into the total electric-field-injected-energy estimate, we obtain 
\bea
\mathcal{E}_{\rm min} \approx \zeta L w^2 n_e e V_{\rm min} \sim \mathcal{O}(\zeta \times 10^{44})-\mathcal{O}(\zeta \times 10^{45})\, \text{J}\,,
\eea
where $\zeta$ is the fraction of runaway electrons, and our computation above establishes that it is not diminutive. Compare $\mathcal{E}_{\rm min}$ with Eqs.~7 through 10 of \cite{2020ApJ...890L..18T}, we see that this energy budget, although oversimplified and ignores possible losses, appears already sufficient to power the observed filament luminosities. Furthermore, $V_{\rm min}$ corresponds to $\gamma \sim \mathcal{O}(10^0)$, and can be increased if we demand a larger $\gamma$. 

In fact, the large value of $\mathcal{E}_{\rm min}$ already seen at $V_{\rm min}$ affords us some freedom to tune down the $\gamma$ factor as compared to the $\gamma \sim 10^3$ assumed by \cite{2020ApJ...890L..18T} in order to land the peak of the synchrotron radiation close to the observation window. In other words, the peak could possibly land at lower frequencies (scales as $\gamma^2$), and there would still be sufficient energy injected into the few (off-peak) GHz windows that we have so far peaked through, to reproduced the observed luminosities. Such flexibility in placing the synchrotron emission peak comes handy when trying to explain the rather significant variations in observed spectral indices. The indices are seen to even differ in sign \cite{1992A&A...264..493L,2022ApJ...925L..18Y}, which could be explained as the employed observation window landing to the left of the peak in some cases, but to the right in others. It is also rather natural that the indices exhibit clear monotonic trends against absolute Galactic latitude (see Fig.~3b of \cite{2022ApJ...925L..18Y}). As the electrons are accelerated gradually across the entire length of the filaments, the high latitudes host less energetic electrons (the electric field points away from the Galactic plane, so electrons are accelerated towards it) whose emission peak lands further to the left of the observation window, thus the window view exhibits a steeper spectrum (in contrast, the regions abutting the peak are flatter). 

\subsection{Axial magnetization \label{sec:axialmag}}
As particle collision is a vital mechanism for randomizing velocities so as to drain energy from bulk into thermal motion, the large mean free paths (see Eq.~\ref{eq:SegVals}) of the wind imply that the plasma is of low Ohmic resistance. The magnetic dominance (see Eq.~\ref{eq:WindBeta}) then further suggests that the (almost\footnote{There may still be anomalous resistance, due to small scale turbulence associated with e.g., shockwaves or magnetic reconnection, as well as inertial resistance due to energy being fed into accelerating charged particles into orderly (rather than random as with Ohmic) motion.}-)ideal magnetohydrodynamics would likely be of a force-free nature\footnote{That is, instead of the magnetic field being dragged around by a dense matter flow that tends to persist in their previous state due to large inertia, the tenuous plasma instead quickly adjusts its motion so as to avoid suffering a net Lorentz force that cannot be balanced by kinetic pressure, which would create a diverging acceleration due to the comparatively paltry inertia.} \cite{1964NASSP..50..389G}, consistent with the filaments settling into a steady-state dominated by an axial magnetic field, that is aligned with the current density ${\bf j}$ so ${\bf j}\times {\bf B}\approx 0$. 

We can speculate on one (of possibly many) pathway for the ascend of this axial field. The magnetic deflection exercise of Sec.~\ref{sec:Current} generates a current, which is subsequently boosted by the acceleration of the electrons during the Z-pinching process. If this current rises above a critical value ($\lambda_{\rm dio} \sim L$, see below), diocotron instability \cite{1950coel.book.....A} would result (magnetic deflection may additionally produce charge separation, cf., footnote \ref{fn:ChargeSep}, also conducive to the development of the diocotron instability). This complication would beget vortical electron motion in the azimuthal direction, and subsequently an axial magnetic field component. Indeed, the $e$-folding length for the buildup of the diocotron instability is given by \cite{1024282,2015ppu..book.....P} 
\bea
\ell_{\rm dio}=\frac{ \lambda_{\rm dio} \pi a B_z V}{I}\,,
\eea
where $\lambda_{\rm dio} \approx (\pi/0.4) a$ is the instability wavelength, $V$ is voltage and $I$ is current. Therefore, the increase of current (and the pinching that reduces the filament width) encourages this instability (smaller $\ell_{\rm dio}$ indicates more rapid development of the instability), while the presence of an axial magnetic field $B_z$ tends to curtail it. Therefore, the diocotron instability will kick off after pinching has progressed to some extent, until a steady-state is reached where $B_z$ grows sufficiently to both quench the instabilities, including diocotron itself (i.e., it is constrained by a negative feedback), and resist further pinching. 

Substituting in some numbers, we see that given the minimal $V_{\rm min}$ from Sec.~\ref{sec:Seg} and $I$ computed in Sec.~\ref{sec:Current}, the diocotron instability would shut down ($\ell_{\rm dio}$ exceeds the size of the system at $L$) and cease enhancing the axial magnetic field, at $B_z \sim \mathcal{O}(10^0)-\mathcal{O}(10^1)\,$mG, overlapping with $B_z^f$. That is, the observed field strength is consistent with the theoretical terminal value predicted with diocotron instability.  Moreover, morphologically, the diocotron instability would further fragment a filament into finer strands that twist around each other\footnote{The self-generated $B^f_z$ can now provide the necessary parity-breaking to sustain consistent  handedness of these strands along the vast length of the filaments, cf., footnote \ref{fn:backbone}. This $B^f_z$ is confined within the spatial expanses of the parent filament however, rather than filling the surrounding environment further out, so the parent filament itself does not coil.}, so the filament becomes a braid when looked closely (see e.g., Fig.~12.7 of \cite{2015ppu..book.....P} for a typical laboratory diocotron snapshot), which appears to match the high resolution scrutinization of the radio filaments (see e.g., Fig.~4 of \cite{2019ApJ...884..170P}).

\section{Conclusion}
In this brief note, we have outlined the rudimentary elements of a proffered mechanism that might be of value to the understanding of the workings of the radio filaments. We have provided some order-of-magnitude impressions to prefatorily investigate whether there are any deficiencies that would immediately invalidate our proposed scheme. Because complicated nonlinear instabilities are both unavoidable and necessary, we are not able to pin down much details, yet the large redundancies of the zeroth order quotations appear reassuring. We therefore hope that this note could serve as the synopsis to a self-consistent thesis, to be fleshed out with more theory work and observations. 

In particular, with luck, more sensitive instruments (e.g., \cite{2017ARep...61..288G,2014SSPMA..44..783W,2023RAA....23i5005X}), soon to come online, might catch filaments currently going through their dimmer formative stages. Such observations would expose additional dynamical details, which when compared with laboratory experiments and perhaps numerical simulations, should provide more definite evidence supporting or refuting our hypothesis. 

Furthermore, since our simple numerology seems to still work well with low electron Lorentz factors, there exists the possibility that some high Galactic latitude filaments are even brighter at ultra-long wavelengths. Observatories (see e.g., \cite{2023ChJSS..43...43C,2021RSPTA.37990560S}) taking advantage of the quiet back side of the Moon to try and paint the Universe at these thus far largely unexplored spectral realms may yield interesting insights. 

\acknowledgements
This work is supported by the National Natural Science Foundation of China grants 12073005 and 12021003, and National Key Research and Development Program of China grant 2023YFC2205801. 
\newpage

%\bibliography{../References}
\bibliography{filaments.bbl}

\appendix
\end{document}